\documentstyle[12pt]{article}
\def\gappeq{\mathrel{\rlap {\raise.5ex\hbox{$>$}}
{\lower.5ex\hbox{$\sim$}}}}

\def\lappeq{\mathrel{\rlap{\raise.5ex\hbox{$<$}}
{\lower.5ex\hbox{$\sim$}}}}

\def\NPB#1#2#3{Nucl. Phys. B {\bf#1} (19#2) #3}
\def\PLB#1#2#3{Phys. Lett. B {\bf#1} (19#2) #3}

\def\PRD#1#2#3{Phys. Rev. D {\bf#1} (19#2) #3}
\def\PRL#1#2#3{Phys. Rev. Lett. {\bf#1} (19#2) #3}

\def\ARNP#1#2#3{Ann. Rev. Nucl. Part. Sci. {\bf#1} (19#2) #3}

\newcommand{\newc}{\newcommand}
\newc{\sm}{Standard Model }
\newc{\smd}{Standard Model}
\newc{\dac}{discrete anomaly cancellation }
\newc{\mup}{``$\mu$'' problem }
\newc{\eps}{\epsilon}
\begin{document}
%\begin{titlepage}

\begin{flushright}
CERN-TH.7412/94\\
HD-HEP-94-30\\
 OUTP-94\\
STUAM-FTUAM.40/94\\
\end{flushright}

\begin{center}
{\bf NEUTRINO MASSES FROM GAUGE SYMMETRIES} \\
\vspace*{0.3cm}
{\bf H. Dreiner$^{1}$,   G.K. Leontaris$^{2}$,
 S. Lola$^{3}$,}\\
{\bf  Graham G. Ross,$^{4,*}$}  and {\bf C. Scheich$^{5}$}  \\
\end{center}
\vspace*{0.1cm}
\begin{center}
\begin{tabular}{l}
$^{1}${\small ETH H\"onggerberg, 8093 Z\"urich, Switzerland}\\
$^{2}${\small CERN, Theory Division, 1211 Geneva 23, Switzerland, and}\\
{\small Physics Department, Ioannina University, Ioannina, Greece}\\
$^{3}${\small Institut f\"{u}r
Theoretische Physik, Univerisit\"at Heidelberg,}\\
{\small Philosophenweg 16, 69120 Heidelberg, Germany }\\
$^{4}${\small Department of Physics,Theoretical Physics, University of
Oxford,}\\
{\small 1 Keble Road, Oxford OX1 3NP}\\
$^{5}${\small Departemento de F\'isica Te\'orica, Universidad Aut\'onoma
de Madrid,}\\ {\small 28049, Madrid, Spain}\\
\end{tabular}
\end{center}

\begin{center}
{\bf ABSTRACT}
\end{center}

\small{
A very simple extension of the \sm to include an Abelian family
symmetry is able to describe the hierarchy of quark and lepton
masses and their mixing angles together with the unification of
gauge couplings. We consider the implications of this model for
neutrino masses and mixing angles and show that they are
determined up to a discrete ambiguity corresponding to the
representation content of the Higgs sector responsible for the
Majorana mass matrix.}

%\vspace*{1cm}
\noindent
\rule[.1in]{14.5cm}{.002in}

\noindent
$^*$ SERC Senior Fellow.
%\vspace*{0.5cm}

\begin{flushleft}
CERN-TH.7412/94\\
HD-HEP-94-30\\
OUTP-94\\
STUAM-FTUAM.40/94\\
August 1994
\end{flushleft}

%\noindent
\thispagestyle{empty}
\setcounter{page}{1}
\vfill\eject

\section{Introduction}

If we are to understand the pattern of fermion  masses and mixing
angles it is necessary to go beyond the \smd. One obvious
possibility is that there is further structure, Grand
Unification, strings etc., at a high scale which determines the
Yukawa couplings responsible for the masses. However attempts to
implement such ideas have to explain why the electroweak breaking
scale is much less than the unification scale. Low energy
supersymmetry \cite{susy} provides a way of protecting such an
        hierarchy of mass scales and if supersymmetry is broken
by a gaugino condensate it can even explain the origin of the
hierarchy.

Within the context of supersymmetry, unification\cite{unif,unif1}
 has
had considerable  success in determining  the parameters of the
\sm \cite{1a}.
The measured values of the  gauge couplings are consistent with
simple unification values with a unification scale of
O($10^{16}GeV$) provided the \sm \ spectrum is extended to that
of the  minimal supersymmetric model (the MSSM). In
addition the pattern (and magnitude) of spontaneous breakdown of
the \sm  follows naturally from the structure of radiative
corrections in the MSSM provided there is some unification of the
supersymmetry breaking masses at the unification
scale\cite{unif,radi}. This simplicity in the parameters of the
(supersymmetric) \sm at high scales appears to extend to the
Yukawa couplings involved in determining the fermion masses. The
measured values of the bottom quark and the $\tau$ lepton masses
are consistent with their equality at the unification
scale \cite{1,1a}.
Further the mixing angles and masses have values
consistent with the appearance of ``texture'' zeros in the mass
matrix\cite{3,4,5,5e}, such texture zeros indicating the
appearance of additional symmetries beyond the \sm. This, and the
hierarchical structure observed in the quark and lepton mass
matrices, strongly suggests the existence of an underlying family
symmetry with breaking characterised by a small parameter,
$\lambda$ \cite{7,IR}. In the limit the symmetry is exact only
the third generation is massive and all mixing angles are zero.
Symmetry breaking terms gradually fill in the mass matrices in
powers of $\lambda$ generating an hierarchy of mass scales and
mixing angles. Thus a broken symmetry can explain not
only the ``texture'' zeros but also the relative magnitude of the
non-zero elements.

It proves remarkably easy to construct such a broken symmetry
giving rise to a realistic mass matrix.  A simple Abelian gauge
family symmetry spontaneously broken close to the Planck scale
generates all the observed structure in the mass matrices. The
structure is consistent with a much larger symmetry
suggesting further unification, namely left-right-symmetry giving
symmetric mass matrices, $SU(2)_R$ symmetry relating up and down
quark couplings and a down- quark lepton symmetry \footnote{Such
an extended symmetry would be consistent with $(SU(3))^3$ or
$E_6$ gauge unification but not with $SU(5)$.}.  In addition,
consistency of this fermion mass structure with  freedom of
anomalies yields the successful prediction
$sin^2(\theta_W)=3/8$ at the unification scale even though the
gauge group is not Grand Unified \cite{luis}.

In this letter we consider the implications of this scheme for
neutrino masses and mixing angles in the case that the minimal
multiplet content of the MSSM is extended to include right-handed
neutrino components (plus the \sm singlet Higgses needed to
generate their masses and to break the extended gauge family
symmetry). Although this extends the multiplet content of the \sm
it represents perhaps the most likely departure from the original
\sm if neutrinos prove to have masses. The introduction of this
sector requires the introduction of a Majorana mass matrix which
has no counterpart in the \sm. Most attempts to deal with this
matrix use very simple but ad hoc assumptions about its
structure\cite{7b}. We will show that, within the context of an
extended family gauge symmetry with symmetric mass matrices, the
structure of the Majorana mass matrix is determined up to a
discrete ambiguity and that the resultant form is different from
the usual assumptions made about its form. We will be
particularly concerned to determine whether the pattern of light
neutrino masses and mixings
resulting from this structure can explain any of the experimental
indications of neutrino mass.

\section{Quark and Charged Lepton Fermion Masses}

We start by reviewing the construction of the model\cite{IR} of
quark and charged lepton masses. The structure of the mass
matrices is determined by a family symmetry, $U(1)_{FD}$, with
in general non-integer charge assignment of the \sm states given
in Table \ref{table:2a}. The need to preserve $SU(2)_L$
invariance requires (left-handed) up and down quarks (leptons)
to have the same charge. This plus the requirement of symmetric
matrices then requires that all quarks (leptons) of the same i-th
generation transform with the same charge $\alpha _i(a_i)$.  The
full anomaly free Abelian group involves an additional family
independent component, $U(1)_{FI}$, and with this freedom we may
make $U(1)_{FD}$ traceless without any loss of generality. Thus
$\alpha_3=-(\alpha_1+\alpha_2)$ and $a_3=-(a_1+a_2)$.
\begin{table}
\begin{center}
\begin{tabular}{|c |cccccccc|}\hline
   &$ Q_i$ & $u^c_i$ &$ d^c_i$ &$ L_i$ & $e^c_i$ & $\nu^c_i$ &
$H_2$ &
$ H_1$   \\
\hline
  $U(1)_{FD}$ & $\alpha _i$ & $\alpha _i$ & $\alpha _i$  & $a_i$
& $a_i$ & $a_i $ & $-2\alpha _1$ &  $-2\alpha _1$
\\
\hline
\end{tabular}
\end{center}
\caption{ $U(1)_{FD}$ symmetries. }
\label{table:2a}
\end{table}

 The $U(1)_{FD}$ charge of the quark-antiquark pair has the  form
\begin{eqnarray}
\left(
\begin{array}{ccc}
-2(\alpha_1+\alpha_2) & -\alpha_1 & -\alpha_2 \\
-\alpha_1 & 2\alpha_2 & \alpha_1 + \alpha_2 \\
-\alpha_2 & \alpha_1 + \alpha_2 & 2\alpha_1
\end{array}
\right)
\label{eq:p}
\end{eqnarray}
This matrix neatly summarises the allowed Yukawa couplings for
a Higgs boson coupling in a definite position. They  should have
charge minus that shown for the relevant position.

For the leptons we have a similar structure of lepton-antilepton
charges
\begin{eqnarray}
\left(
\begin{array}{ccc}
-2(a_1+a_2) & -a_1 & -a_2 \\
-a_1 & 2a_2 & a_1 + a_2 \\
-a_2 & a_1 + a_2 & 2a_1
\end{array}
\right)
\label{eq:pl}
\end{eqnarray}
If the light Higgs, $H_{2}$, $H_{1}$, responsible for the up and
down quark masses respectively have $U(1)$ charge so that only
the (3,3) renormalisable Yukawa coupling to $H_{2}$, $H_{1}$ is
allowed, only the (3,3) element of the associated mass matrix
will be
non-zero as desired. The remaining entries are generated when the
$U(1)$ symmetry is broken. We assume this breaking is spontaneous
via \sm singlet fields,
$\theta,\; \bar{\theta}$, with $U(1)_{FD}$ charge -1, +1
respectively, which
acquire equal vacuum expectation values (vevs) along a ``D-flat''
direction\footnote{The spontaneous breaking of gauge symmetries
at high scales in  supersymmetric theories must proceed along
such flat directions to avoid large vacuum energy contributions
from D-terms, giving $<\theta>=<\bar{\theta}>$.}.  After this
breaking all entries in the mass matrix become non-zero. For
example, the (3,2) entry in the up quark mass matrix appears at
$O(\eps^{\mid\alpha_2-\alpha_1 \mid} )$ because  U(1) charge
conservation allows only a coupling       $c^c t H_2(\theta
/M_2)^{\alpha_2-\alpha_1}, \;  \alpha_2>\alpha_1$ or   $c^ct
H_2(\bar{\theta} /M_2)^{\alpha_1-\alpha_2},\; \alpha_1>\alpha_2$
and we have defined
$\epsilon=(<\theta>/M_2)$ where $M_2$ is the unification mass
scale  which governs the higher dimension operators. As discussed
in reference\cite{IR} one may expect a different scale, $M_{1}$,
for the down quark mass matrices (it corresponds to mixing in the
$H_{2}$, $H_{1}$ sector with $M_{2}$, $M_{1}$ the masses of heavy
$H_{2}$, $H_{1}$ fields). Thus we arrive at mass matrices of the
form (suppressing unknown Yukawa couplings and their phases which
are all expected to be of order 1)
\begin{eqnarray}
\frac{M_u}{m_t}\approx \left(
\begin{array}{ccc}
\epsilon^{\mid 2+6a \mid } &
\epsilon^{\mid 3a \mid } &
\epsilon^{\mid 1+3a\mid }
\\
\epsilon^{\mid 3a \mid } &
\epsilon^{ 2 } &
\epsilon^{ 1 } \\
\epsilon^{\mid 1+3a \mid } &
\epsilon^{1 } & 1
\end{array}
\right)
\label{eq:mu0}
\end{eqnarray}
\begin{equation}
\frac{M_d}{m_b}\approx \left (
\begin{array}{ccc}
\bar{\epsilon}^{\mid 2+6a \mid } &
\bar{\epsilon}^{\mid 3a \mid } &
\bar{\epsilon}^{\mid 1+3a \mid } \\
\bar{\epsilon}^{\mid 3a \mid } &
\bar{\epsilon}^{ 2 } &
\bar{\epsilon}^{ 1 } \\
\bar{\epsilon}^{\mid 1+3a \mid } &
\bar{\epsilon}^{1} & 1
\end{array}
\right)
\label{eq:massu}
\end{equation}
where $\bar{\epsilon} = (\frac{<\theta >}{M_1})^{|\alpha_2-
\alpha_1|}$,
$\epsilon=(\frac{<\theta >}{M_2})^{|\alpha_2-\alpha_1|}$ and
$a=\alpha_1/(\alpha_2-\alpha_1)$.
With $a=1$ the mass matrices are in excellent agreement with the
measured values. We also point out that to a good approximation
we have the relation \cite{IR}
\begin{equation}
\eps=\bar{\eps}^2
\label{eq:eps}
\end{equation}
which also implies that $M_2>M_1$. This relation will be very
helpful  below when determining the structure of the neutrino
spectrum.

The charged lepton mass matrix may similarly be determined.
Requiring the good relation $m_b=m_{\tau}$ at unification sets
$\alpha_1=a_1$ and we get
\begin{equation}
\frac{M_l}{m_{\tau}}\approx \left (
\begin{array}{ccc}
\bar{\epsilon}^{\mid 2+6a-2b \mid } &
\bar{\epsilon}^{\mid 3a \mid } &
\bar{\epsilon}^{\mid 1+3a-b \mid } \\
\bar{\epsilon}^{\mid 3a \mid } &
\bar{\epsilon}^{ \mid 2(1-b) \mid } &
\bar{\epsilon}^{ \mid 1 -b \mid} \\
\bar{\epsilon}^{\mid 1+3a-b \mid } &
\bar{\epsilon}^{\mid 1-b \mid} &1
\end{array}
\right)
\label{eq:7}
\end{equation}
where $b=(\alpha_2-a_2)/(\alpha_2-\alpha_1)$. We will consider
two choices for $b$ which give reasonable lepton masses.

For $b=0$ the lepton charges are the same as the down quark
sector, and so the structure of the down quark and lepton mass
matrices are identical. In order to explain the detailed
difference between down quark and lepton masses it is necessary
in this case to assume that the constants of proportionality
determined by Yukawa couplings which we have so far taken to be
equal (and of O(1)) differ sightly for the lepton case. A factor
3 in the (2,2) entry is sufficient to give excellent charged
lepton masses.

An alternative which does not rely on different Yukawa couplings
is to choose $b$ half integral. In this case the form for the
lepton mass matrix, eq(\ref{eq:7}) must be modified and has the
form
\begin{equation}
M_l = \left (
\begin{array}{ccc}
\bar{\epsilon}^{\mid 2+6a-2b \mid } &
\bar{\epsilon}^{\mid 3a \mid } &
0 \\
\bar{\epsilon}^{\mid 3a \mid } &
\bar{\epsilon}^{ \mid 2(1-b) \mid } &
0 \\
0 & 0 & 1
\end{array}
\right)
\label{eq:7l}
\end{equation}
The zeros in this mass matrix result because  there is now a
residual $Z_2$ discrete gauge symmetry after $U(1)$ breaking by
which the  electron
and muon fields get transformed by a factor $(-1)$. For $a=1$,
$b=1/2$ we get excellent agreement for the charged lepton masses;
for this choice one gets the following relations at laboratory
energies
\begin{equation}
\begin{array}{ccc}
m_{\mu}&\approx& \frac{m_s}{3\bar{\epsilon}} \\
m_e&\approx &\frac{m_d \bar{\epsilon}}{3}
\end{array}
\label{eq:mue}
\end{equation}
where the factor 3 is due to the renormalisation group
corrections
found when continuing from the Grand Unification scale to the
laboratory scale.  This is in good agreement with the
experimentally measured values for the value
$\bar{\epsilon}=0.23$ needed to fit the down quark masses and
mixing angles. Using Eq.(\ref{eq:eps}) this in turn implies that
$\eps=0.053$.

\section{Neutrino Mass Matrices}

We turn now to a discussion of the implications of this broken
symmetry for neutrino masses in the case we add three generations
of right-handed neutrinos.  Note first that $SU(2)_L$ fixes the
$U(1)_{FD}$ charge of the left-handed neutrino states to be the
same as the charged leptons. The left- right- symmetry then fixes
the charges of the right-handed neutrinos as given in Table
\ref{table:2a}. Thus we have no freedom in assigning right-handed
neutrino charges and as a result the neutrino Dirac mass is also
fixed to be
\begin{equation}
\frac{M^D_{\nu_R}}{m_{\nu_{\tau}}}\approx \left (
\begin{array}{ccc}
{\epsilon}^{\mid 2+6a-2b \mid } &
{\epsilon}^{\mid 3a \mid } &
{\epsilon}^{\mid 1+3a-b \mid } \\
{\epsilon}^{\mid 3a \mid } &
{\epsilon}^{ \mid 2(1-b) \mid } &
{\epsilon}^{ \mid 1 -b \mid} \\
{\epsilon}^{\mid 1+3a-b \mid } &
{\epsilon}^{\mid 1-b \mid} & 1
\end{array}
\right)
\label{eq:nud}
\end{equation}
for $b$ integer or
\begin{equation}
\frac{M^D_{\nu}}{m_{\nu_{\tau}}}\approx \left (
\begin{array}{ccc}
{\epsilon}^{\mid 2+6a-2b \mid } &
{\epsilon}^{\mid 3a \mid } &
0 \\
{\epsilon}^{\mid 3a \mid } &
{\epsilon}^{ \mid 2(1-b) \mid } &
0 \\
0 & 0 & 1
\end{array}
\right)
\label{eq:nudp}
\end{equation}
for $b$ half integer. Thus the Dirac neutrino mass matrices are
related  in  scale to the up-quark mass matrices, similar to
models based on Grand Unified Theories.

Of course the mass matrix structure of neutrinos is more
complicated due to the possibility of Majorana masses for the
right-handed components\footnote{For the left-handed
components these do not appear for the usual reason, namely the
absence of $\Delta I_W=1$ mass terms coming from weak isospin 0
($I_W=0$) Higgs fields.}. The right handed fields on the other
hand  can get Majorana masses from a term of the form
$\nu_R\nu_R\Sigma$  where $\Sigma$ is a $SU(3)\otimes
SU(2)\otimes U(1)$ invariant Higgs scalar field with $I_W=0$ and
$\nu_R$ is a right-handed neutrino.  In many
models $\Sigma$\cite{su3} is a
combination of scalar fields $\Sigma = \tilde{\bar{\nu}}_R
\tilde{\bar{\nu}}_R$ where $\tilde{\bar{\nu}}_R$ is the scalar
component of a right-handed antineutrino supermultiplet:
${\bar\nu}_R$.  For  definiteness we mostly focus on this model
but  our main results  do not  depend on this choice.

The structure of the resulting Majorana mass matrix depends on
the $U(1)_{FD}$ invariant combinations contributing. Clearly the
right-handed neutrino bilinears have the $U(1)_{FD}$ charge
structure of eq(\ref{eq:pl}). Thus the Majorana mass matrix,
$M^M_{\nu}$, depends entirely on the $\Sigma$ $U(1)_{FD}$ charge.
The possible choices for it will give a discrete spectrum of
possible forms for the Majorana mass.

For example if, in the absence of $U(1)_{FD}$ symmetry breaking
the $\Sigma$ charge is the same as the $H_{1,2}$ doublet Higgs
charges, only the (3,3) element of $M_{\nu}$ will be non-zero.
Allowing for $U(1)_{FD}$ breaking by $<\theta>$ the remaining
elements  in the Majorana mass matrix will be
generated in an analogous way to the generation of the Dirac mass
matrices.\footnote{$<\Sigma>$ is significantly below the Planck
scale and thus $<\theta>$ dominates the $U(1)_{FD}$ breaking.}
The important question is what is the appropriate
expansion parameter? It will be given by
$\eta=\frac{<\Theta >}{M_3}$, where $M_3$ is the scale
determining the mixing in the $I_W=0$ sector. In the case of
$H_{1}$,$H_{2}$ the scale was identified with the massive
$H_{1}$,$H_{2}$ fields which mix with the light $H_{1}$,$H_{2}$
fields when $<\theta >$ is non-zero. Likewise, in the $I_W=0$
Higgs sector $M_3$ should be identified with the massive $\Sigma$
modes. For the case that $\Sigma$ is the bilinear $\Sigma =
\tilde{\bar{\nu}}_R \tilde{\bar{\nu}}_R$ it is the mixing of the
${\bar{\nu}}_R$ fields that generates the mixing and hence the
mass of the intermediate ${\bar{\nu}}_R$ fields that is relevant.

What is the expectation for such masses? The pattern of masses
generated here relies on the mixing of light and heavy Higgs
fields. Consider a string compactification which in addition to
$H_1$ and $H_2$, leaves additional Higgs multiplets
$H_{1,2}^{a,b...},\bar{H}_{1,2}^{a,b...}$ light. The pairs of
Higgs fields in conjugate representations can have gauge
invariant masses and may be expected to become massive if there
is any stage of spontaneous symmetry breaking below the
compactification scale with a common mass, $M$, where $M=<\Phi>$
and $\Phi$ is a gauge invariant Higgs scalar field (or a gauge
invariant combination of Higgs fields). There may be further
sources of Higgs mass. As we have stressed our left- right-
symmetry essentially requires an extension of the gauge symmetry
to $SU(2)_L\otimes SU(2)_R$ at high scales. This will be broken
by a right-handed sneutrino vev in which case the mass degeneracy
of the $H_1$ and $H_2$ pair which transform as a $(1/2,1/2)$
representation under $SU(2)_L\otimes SU(2)_R$ can be split via
the coupling $<\tilde \nu_R>H_2H_x$ where $H_x$ transforms as
$(1/2,0)$. Such a contribution will generate $M_2\approx
<\tilde \nu_R>$, $M_1\approx M$
agreeing well with Eq.(\ref{eq:eps}) and
the phenomenological findings of \cite{IR}.

Turning to the expectation for $\nu_R$ masses we may expect the
${\bar{\nu}}_R$ fields to aquire mass of $O(M_1)$ via a
$\Phi\nu{\bar\nu}$ coupling. These fields may also acquire a mass
via the term $<\tilde \nu_R>^2(\bar{\nu}_R)^2$ giving a mass
$<\tilde \nu_R>^2/M_{Pl}$. Given that $<\tilde \nu_R>\approx M_2$
then $\epsilon=\bar\epsilon^2$ from
Eq.(\ref{eq:eps}) implies that  $<\tilde \nu_R>^2/M_{Pl} =M_1.
(M_2/M_{Pl})/\bar{\epsilon}$. To preserve the successful
relations between up and down quarks we require that
non -- renormalisable contributions to the up quark masses should
be less than that coming from the mixing with the heavy $H_2$
fields and so $M_2/M_{Pl}<<1$.  Thus these terms are not expected
to spoil our estimate that $m_{\nu_R}\approx M_1$.

This implies that the appropriate expansion parameter for the
Majorana  mass matrix is the same as that for the {\it down}
quarks and charged  leptons, namely $\eta\approx\bar{\epsilon}$.
In a specific model this  relation can be determined exactly. In
general we only expect this to  hold up to factors of order one.

We may now compute the patterns of Majorana mass for the
different possible choices of $\Sigma$ charge. These are given
in Tables \ref{table:maj} and \ref{table:majp} together with the
mass eigenvalues. We have made use of Eq.(\ref{eq:eps}).
Note the result is quite different from the
usual ansatz which assumes it is family independent. In fact none
of the possibilities of Table \ref{table:maj} give a family
independent structure for the Majorana mass. This is perhaps the
most important lesson to be derived from this model; if
symmetries are responsible for the hierarchical pattern of
fermion masses they are very likely to generate an hierarchical
pattern for the Majorana mass matrix.
Moreover, as may be seen from  Table \ref{table:maj},
in most of the cases the lighter  $M_{\nu_R}$ eigenstate is
suppressed by several powers of $\bar \epsilon$ compared to
the heavier one.
The implications for the
phenomenology of neutrino masses is quite different as we shall
now discuss.

Of relevance to low energy phenomena is the pattern of light
neutrino masses and mixing angles. To compute these we must
determine the light neutrino mass matrix, $M^{eff}_{\nu}$, coming
from the ``see-saw'' mechanism.  It is given by
\begin{equation}
M^{eff}_{\nu}=M^D_{\nu}\cdot (M^M_{\nu_R})^{-1}\cdot M^D_{\nu}
\label{eq:meff}
\end{equation}
Using the two forms for the neutrino Dirac matrices given in
eqs(\ref{eq:nud}) and (\ref{eq:nudp}) together with the
appropriate possibilities for the Majorana mass matrix, Tables
\ref{table:maj} and \ref{table:majp} respectively, we may compute
the masses and mixing angles for the light neutrinos. These are
given in Tables \ref{table:meff} and \ref{table:meffp}
respectively. The use of Eq.(\ref{eq:eps}) greatly simplifies the
results.

\begin{table}
\centering
\begin{tabular}{|c|c|c|c|} \hline
Case & $Q_{\Sigma}/a_1+2$ & ${\bf M_{\nu_R}}$ & $M_{\nu_R}^{Diag}$
\\ \hline
1 & 0 & $\left(
\begin{array}{ccc}

 \bar{\epsilon}^8 & \bar{\epsilon}^3 & \bar{\epsilon}^4 \\
\bar{\epsilon}^3 & \bar{\epsilon}^2 & \bar{\epsilon} \\
 \bar{\epsilon}^4 & \bar{\epsilon} & 1
\end{array}
\right)$ &
$\left(
\begin{array}{ccc}

 \bar{\epsilon}^4 &  &  \\
  & \bar{\epsilon}^2 &  \\
  &  & 1
\end{array}
\right)$
\\ \hline

2 & -1 & $\left(
\begin{array}{ccc}
\bar{\epsilon}^9 & \bar{\epsilon}^4 & \bar{\epsilon}^5 \\
\bar{\epsilon}^4 & \bar{\epsilon} & 1 \\
\bar{\epsilon}^5 & 1 & \bar{\epsilon}
\end{array}
\right)$ &
$\left(
\begin{array}{ccc}
\bar{\epsilon}^9 &  &  \\
 & -1+\bar{\epsilon} &  \\
 &  & 1+\bar{\epsilon}
\end{array}
\right)$
\\ \hline

3 & -2 & $\left(
\begin{array}{ccc}
\bar{\epsilon}^{10} & \bar{\epsilon}^5 & \bar{\epsilon}^6 \\
\bar{\epsilon}^5 & 1 & \bar{\epsilon} \\
\bar{\epsilon}^6 & \bar{\epsilon} & \bar{\epsilon}^2
\end{array}
\right)$
&
$\left(
\begin{array}{ccc}

\bar{\epsilon}^{10} &  &  \\
  & 1 &  \\
  &  & \bar{\epsilon}^2
\end{array}
\right)$
\\ \hline

4 & 2& $\left(
\begin{array}{ccc}
\bar{\epsilon}^5 & 1 & \bar{\epsilon} \\
1 & \bar{\epsilon}^5 & \bar{\epsilon}^4 \\
\bar{\epsilon} & \bar{\epsilon}^4 & \bar{\epsilon}^3
\end{array}
\right)$ &
$\left(
\begin{array}{ccc}

1-\bar{\epsilon}^2 &  &  \\
  & -1+\bar{\epsilon}^2 &  \\
  &  & \bar{\epsilon}^3
\end{array}
\right)$

\\ \hline

5 & 4 & $\left(
\begin{array}{ccc}
\bar{\epsilon}^4 & \bar{\epsilon} & 1 \\
\bar{\epsilon} & \bar{\epsilon}^6 & \bar{\epsilon}^5 \\
1 & \bar{\epsilon}^5 & \bar{\epsilon}^4
\end{array}
\right)$ &
$\left(
\begin{array}{ccc}

1-\bar{\epsilon}^2 &  &  \\
  & \bar{\epsilon}^6 &  \\
  &  & -1+\bar{\epsilon}^2
\end{array}
\right)$

\\ \hline

6 & 8 & $\left(
\begin{array}{ccc}
1 & \bar{\epsilon}^5 & \bar{\epsilon}^4 \\
\bar{\epsilon}^5 & \bar{\epsilon}^{10} & \bar{\epsilon}^9 \\
\bar{\epsilon}^4 & \bar{\epsilon}^9 & \bar{\epsilon}^8
\end{array}
\right)$ &
$\left(
\begin{array}{ccc}

 1 &  &  \\
  & \bar{\epsilon}^{10} &  \\
  &  &  \bar{\epsilon}^{8}
\end{array}
\right)$

\\ \hline
\end{tabular}
\caption{Majorana mass matrix, $M_{\nu_R}$, for the right-handed
neutrinos for a range of choices of $Q_{\Sigma}$, the  $U(1)$
charge of the $I_W^R=2$ effective Higgs field. These cases
correspond to $b=0$ and $a=1$ in eq(6). Note that we have
suppressed all coefficients of O(1). For example in case 1 the
$(2\otimes2)$ submatrix comprising the second and third rows and
columns does {\it not} have vanishing determinant for random
choices of these coefficients and the mass eigenvalues quoted
refer to such generic cases. }
\label{table:maj}
\end{table}

\begin{table}
\centering
\begin{tabular}{|c|c|c|c|} \hline
Case & $Q_{\Sigma}/a_1+2$ & ${\bf M_{\nu_R}}$ & $M_{\nu_R}^{Diag}$
\\ \hline
7 & 0 & $\left(
\begin{array}{ccc}

 \bar \epsilon^7 & \bar \epsilon^3 & 0 \\
 \bar \epsilon^3 & \bar \epsilon & 0 \\
 0 & 0 & 1
\end{array}
\right)$ &
$\left(
\begin{array}{ccc}

 \bar \epsilon^5 &  &  \\
  & \bar \epsilon &  \\
  &  & 1
\end{array}
\right)$
\\ \hline

8 & -1 & $\left(
\begin{array}{ccc}
\bar \epsilon^8 & \bar \epsilon^4 & 0 \\
\bar \epsilon^4 & 1 & 0 \\
0 & 0 & \bar \epsilon
\end{array}
\right)$ &
$\left(
\begin{array}{ccc}
 \bar \epsilon^8 &  &  \\
 & 1 &  \\
 &  & \bar \epsilon
\end{array}
\right)$
\\ \hline

9 & 3 & $\left(
\begin{array}{ccc}
\bar \epsilon^{4} & 1 & 0 \\
1 & \bar \epsilon^4 & 0 \\
0 & 0 & \bar \epsilon^3
\end{array}
\right)$
&
$\left(
\begin{array}{ccc}

-1 &  &  \\
  & 1 &  \\
  &  & \bar \epsilon^3
\end{array}
\right)$
\\ \hline

10 & 7& $\left(
\begin{array}{ccc}
1 & \bar \epsilon^4 & 0 \\
\bar \epsilon^4 & \bar \epsilon^8 & 0 \\
0 & 0 & \bar \epsilon^7
\end{array}
\right)$ &
$\left(
\begin{array}{ccc}

1 &  &  \\
  &  \bar \epsilon^8 &  \\
  &  & \bar \epsilon^7
\end{array}
\right)$

\\ \hline
\end{tabular}
\caption{Majorana mass matrix, $M^M_{\nu_R}$, for the right-handed
neutrinos for a range of choices of $Q_{\Sigma}$, the  $U(1)$
charge of the $I_W^R=2$ effective Higgs field. These cases
correspond to $b=1/2, a=1$ in eq(6). }
\label{table:majp}
\end{table}

\begin{table}
\centering
\begin{tabular}{|c|c|c|} \hline
Case & $M^{eff,Diag}_{\nu}$ & $R_{\nu}^{eff}$ \\ \hline
1 &
$\left(
\begin{array}{ccc}

 \bar{\epsilon}^{11} &  &  \\
  & \bar{\epsilon}^7 &  \\
  &  & 1
\end{array}
\right)$ &
 $\left(
\begin{array}{ccc}

 1 & \bar{\epsilon}^3 & -\bar{\epsilon}^5 \\
 -\bar{\epsilon}^3 & 1 & \bar{\epsilon}^2 \\
 \bar{\epsilon}^5 &-\bar{\epsilon}^2 & 1
\end{array}
\right)$
\\ \hline
2 &
$\left(
\begin{array}{ccc}

 \bar{\epsilon}^{11} &  &  \\
  &\bar{\epsilon}^5 &  \\
  &  & \bar{\epsilon}^{-1}
\end{array}
\right)$ &
 $\left(
\begin{array}{ccc}

 1 & \bar{\epsilon}^3 & -\bar{\epsilon}^7 \\
 -\bar{\epsilon}^3 & 1 & \bar{\epsilon}^2 \\
 \bar{\epsilon}^7 & -\bar{\epsilon}^2 & 1
\end{array}
\right)$
\\ \hline
3 &
$\left(
\begin{array}{ccc}

  \bar{\epsilon}^{12} &  &  \\
  &  \bar{\epsilon}^2 &  \\
  &  &\bar{\epsilon}^{-2}
\end{array}
\right)$ &
 $\left(
\begin{array}{ccc}

 1 & \bar{\epsilon}^5 & -\bar{\epsilon}^7 \\
 -\bar{\epsilon}^5 & 1 &  {\bar{\epsilon}^{2}}  \\
 \bar{\epsilon}^7 & {\bar{\epsilon}^{2}}  &1
\end{array}
\right)$
\\ \hline
4 &
$\left(
\begin{array}{ccc}

 \bar{\epsilon}^{15} &  &  \\
  & \bar{\epsilon}^{9} &  \\
  &  & \bar{\epsilon}^{-3}
\end{array}
\right)$ &
 $\left(
\begin{array}{ccc}

 1 & \bar{\epsilon}^3 & -\bar{\epsilon}^7 \\
 -\bar{\epsilon}^3 & 1 & \bar{\epsilon}^2 \\
 \bar{\epsilon}^7 & -\bar{\epsilon}^2 & 1
\end{array}
\right)$
\\ \hline
5 &
$\left(
\begin{array}{ccc}

 \bar{\epsilon}^{13} &  &  \\
  & \bar{\epsilon}^9 &  \\
  &  & \bar{\epsilon}^{-4}
\end{array}
\right)$ &
 $\left(
\begin{array}{ccc}

 1 & \bar{\epsilon}^3 & -\bar{\epsilon}^5 \\
- \bar{\epsilon}^3 & 1 & \bar{\epsilon}^2 \\
 \bar{\epsilon}^5 & -\bar{\epsilon}^2 & 1
\end{array}
\right)$
\\ \hline
6 &
$\left(
\begin{array}{ccc}

  \bar{\epsilon}^{16} &  &  \\
  & \bar{\epsilon}^{-2} &  \\
  &  & \bar{\epsilon}^{-8}
\end{array}
\right)$ &
 $\left(
\begin{array}{ccc}

 1 & \bar{\epsilon}^3 & -\bar{\epsilon}^5 \\
 -\bar{\epsilon}^3 & 1 &\bar{\epsilon}^2 \\
 \bar{\epsilon}^5 & -\bar{\epsilon}^2 & 1
\end{array}
\right)$

\\ \hline
\end{tabular}
\caption{Masses and mixing angles for the light neutrino
components.}
\label{table:meff}
\end{table}

\begin{table}
\centering
\begin{tabular}{|c|c|c|} \hline
Case & $M^{eff,Diag}_{\nu}$ & $R_{\nu}^{eff}$ \\ \hline
7 &
$\left(
\begin{array}{ccc}

 \bar{\epsilon}^{13} &  &  \\
  & \bar{\epsilon}^5 &  \\
  &  & 1
\end{array}
\right)$ &
 $\left(
\begin{array}{ccc}

 1 & \bar{\epsilon}^4 & 0 \\
 -\bar{\epsilon}^4 & 1 & 0 \\
 0 & 0 & 1
\end{array}
\right)$
\\ \hline
8 &
$\left(
\begin{array}{ccc}

 \bar{\epsilon}^{12} &  &  \\
  & \bar{\epsilon}^4 &  \\
  &  & \bar{\epsilon}^{-1}
\end{array}
\right)$ &
 $\left(
\begin{array}{ccc}

 1 & \bar{\epsilon}^4 & 0 \\
 -\bar{\epsilon}^4 & 1 & 0 \\
0 & 0 & 1
\end{array}
\right)$

\\ \hline
9 &
$\left(
\begin{array}{ccc}

 \bar{\epsilon}^{16} &  &  \\
  & \bar{\epsilon}^8 &  \\
  &  & \bar{\epsilon}^{-3}
\end{array}
\right)$ &
 $\left(
\begin{array}{ccc}

 1 & \bar{\epsilon}^4 & 0 \\
 -\bar{\epsilon}^4 & 1 & 0 \\
 0 & 0 & 1
\end{array}
\right)$
\\ \hline
10 &
$\left(
\begin{array}{ccc}

 \bar{\epsilon}^{20} &  &  \\
  & \bar{\epsilon}^{-4} &  \\
  &  & \bar{\epsilon}^{-7}
\end{array}
\right)$ &
 $\left(
\begin{array}{ccc}

 1 & \bar{\epsilon}^4 & 0 \\
- \bar{\epsilon}^4 & 1 & 0 \\
 0 & 0 & 1
\end{array}
\right)$
\\ \hline
\end{tabular}
\caption{Masses and mixing angles for the light neutrino
components.}
\label{table:meffp}
\end{table}

As mentioned above, in no case does the light Majorana mass matrix
 have
degenerate eigenvalues, the most common assumption made for its
structure. The reason is that the gauge symmetry charges of the
right-handed neutrinos force the mass matrix entries to be of
different orders in powers of the expansion parameter
$\epsilon$. The only way a degenerate set of eigenvalues can
occur in this case is if two components are coupled through an
off diagonal mass term as in cases 2, 4, 5 and 9. As may be seen
this only leads to two of the three eigenvalues being
approximately equal. The implications of this for the light
neutrinos is clear from Tables \ref{table:meff} and
\ref{table:meffp}. In no case are the mass eigenvalues simply
given in terms of the square of the up quark masses divided by
a common Majorana mass (the usual ansatz, giving the ratio
$1:\bar \epsilon^2:\bar\epsilon^4$). Rather the effect of
structure  in the Majorana masses is to allow for a greater
spread in masses. As may be seen from these Tables there  is no
example of closely degenerate pairs of neutrino masses; clearly
this is of
considerable significance for neutrino oscillation phenomenology.
In particular all three hints for neutrino masses
(solar neutrino problem, atmospheric neutrino problem, and the
structure formation problem, requiring a one third component of
hot dark matter \cite{hotdark}) taken at face value  can only be
resolved in  agreement with collider bounds (without the use of
additional sterile neutrinos) for three nearly degenerate
neutrinos of approximate mass $2$-$3\,eV$ \cite{psm}.  This
can not be accomodated in our scheme without fine tuning.
However, as we discuss below, we can find simultaneous solutions
to the solar  neutrino problem and the neutrino mass needed for
structure formation.

\section{Neutrino Mixing Angles}

The second point of interest are the predictions for mixing
angles in the light neutrino  sector. It may be seen that the
consequence of the gauge symmetry is to generate mixing angles
which are small, given by some power of the expansion parameter
$\bar \epsilon$. Of course the mixing angles of relevance to
experiment are the combination of charged lepton mixing angles
and neutrino mixing angles given by $R^{eff}_{total}=
({R^{eff}})^{-1}_{\nu}R_{L}$,  where
\begin{equation}
R_{L}=\left(
\begin{array}{ccc}
 1 &  \delta_{e\mu} &  O(\bar \epsilon^4) \\
-\delta_{e\mu} & 1 & O(\bar \epsilon) \\
-O(\bar \epsilon^4) &-O(\bar \epsilon) & 1
\end{array}
\right)
\label{eq:lmix}
\end{equation}
for the charged lepton mass matrix of eq(\ref{eq:7}) with $b=0$
and $a=1$. Here, due to the (1,1) approximate ``texture'' zero
$\delta_{e\mu}=\sqrt{m_e/m_{\mu}}$.
For the charged lepton mass matrix of eq(\ref{eq:7l}) with
$b=1/2$, $a=1$ we have

\begin{equation}
R_{L}=\left(
\begin{array}{ccc}
 1 &  \delta_{e\mu} &  0 \\
-\delta_{e\mu} & 1 & 0 \\
0 & 0 & 1
\end{array}
\right)
\label{eq:lmixp}
\end{equation}

{}From Table \ref{table:meff} we may easily determine that the
only
effect of the neutrino mixing angles for cases  $(1-6)$
is in the (1,3) element of $R^{eff}_{total}$ where the term of
order $\epsilon^4$ in $R_L$ becomes of order $\epsilon^3$ in
$R^{eff}_{total}$. The remaining elements are the same as in
$R_L$, eq(\ref{eq:lmix}) or eq(\ref{eq:lmixp}), because of the
smallness of the
neutrino mixing angles.

The situation for the cases $(7 - 10)$  is simpler due to the
residual $Z_2$ symmetry. For them in all cases the neutrino
mixing angles are so small that $R^{eff}_{total}\approx R_L$. In
this case there is no mixing $\nu_\mu-\nu_\tau$. These cases
could be excluded by a positive finding of the CHORUS and
NOMAD experiments \cite{chorus}.

\section{Phenomenology}

Let us now discuss the implications for neutrino phenomenology
following from these mass structures. Although we have determined
the relative magnitude of the $M_{\nu}^{M}$ and $M^{eff}_{\nu}$,
we are of course interested in the expectation for their absolute
magnitudes. This depends on the origin of the $I_W^R=2$ effective
Higgs field, $\Sigma$.  If $\Sigma = \tilde{\bar{\nu}}_R
\tilde{\bar{\nu}}_R$  the Majorana masses of Tables
\ref{table:maj} and \ref{table:majp} are given in units
$<\tilde{\bar{\nu}}_R><\tilde{\bar{\nu}}_R>/M$ where M is the
mass scale governing the appearance of higher dimension
operators, typically the string scale or $M_{Planck}$. Given the
success of the unification of gauge couplings at a scale of
$O(10^{16}GeV)$ it is reasonable to choose
$<\tilde{\nu}_R>=O(10^{16}GeV)$ leading to the scale $10^{13}-
10^{14}GeV$ for the Majorana mass scale. This in turn implies the
unit of mass for the  light neutrinos masses given in Tables
(\ref{table:meff},\ref{table:meffp}) is
$(4-0.4)eV$ for a top quark of $O(200)GeV$. This means that it
is quite reasonable for the heaviest neutrino to have a mass of
the magnitude needed for structure formation (i.e. of $O(10eV)$)
\cite{hotdark}.

Remarkably, as we will now discuss, for a subset of the solutions
the light neutrinos have masses and mixing angles of the
magnitude needed to explain solar neutrino oscillations.
The experimental data on solar neutrino observations may be
explained if  the masses and mixing angles of the neutrinos fall
in the following ranges\cite{solar}:

a) The small mixing angle solution for the MSW effect requires
\begin{eqnarray}
\delta m^2_{\nu_e\nu_{\mu}}&\approx &(0.6-1.2)\times 10^{-5}eV^2\\
sin^22\theta_{\mu e}&\approx &(0.6-1.4)\times 10^{-2}
\end{eqnarray}

b) Vacuum oscillations can solve the solar neutrino if
\begin{eqnarray}
\delta m^2_{\nu_e\nu_{\mu}}&\approx &(0.5-1.1)\times 10^{-10}eV^2\\
sin^22\theta_{\mu e}&\ge  &0.75
\end{eqnarray}

Consider the small mixing angle solution. If the lightest two
neutrinos are
to have masses in the range needed for the MSW explanation of the
solar neutrino deficit we need a suppression factor of
$O(10^{-3})$ relative to the heaviest neutrino corresponding to
$O(\bar \epsilon^4)$ for $\bar \epsilon\approx 0.23$. We see this
is true for case 3 and is marginally consistent too
with cases 7, 8 and 10. We may go further and consider whether
the
expectation for the mixing angle between the two lightest
generations is in the range needed by the MSW effect. The mixing
angle of relevance to the solar neutrino problem is
$(R^{eff}_{total})_{\mu e}$. In most of the cases presented in
the tables the light neutrino mixing matrices exhibit negligible
$\nu_{\mu}$ - $\nu_e$ mixing and $(R^{eff}_{total})_{\mu e}$
is mainly determined by the charged lepton mixing angle
$\sqrt{\frac{m_e}{m_{\mu}}}$. Although of the correct order of
magnitude, in detail this mixing is larger than that required to
solve the solar neutrino problem. However in cases 1, 2, 4, 5 and
6
the contribution to $\nu_{\mu}$ - $\nu_e$ mixing from the
neutrino
sector is not entirely negligible and we have
\begin{equation}
(R^{eff}_{total})_{\mu e}\approx \sqrt{\frac{m_e}{m_{\mu}}}-
O(\bar{\epsilon}^3)
\label{eq:ll}
\end{equation}
For $\bar{\epsilon}=0.23$ and taking the constant of
proportionality to be unity, one gets $\sin^22\theta_{e\mu}\ge
1.3 \; 10^{-2}$ which is within the limits of the mixing required
to explain the solar neutrino problem.

The vacuum solution to the solar neutrino problem as well as the
neutrino oscillation solution to the atmospheric neutrino problem
require large mixing angles of $O(1)$. As may be seen from the
Tables this is not to be expected in the $U(1)$ family symmetry
presented here. Large mixing angles {\it can} be obtained but
only by fine tuning of the Yukawa couplings. As we have stressed
although we cannot rule out such a possibility we consider it
unlikely as no symmetry requires such fine tuning.

Finally we comment on how one may choose between  the ten
different cases discussed above. As we have stressed they
correspond to possible choices for the $\Sigma$ charge. Is this
constrained? In \cite{IR} the constraints of anomaly freedom were
used to limit the possible
fermion mass matrix structures coming from the $U(1)_{FD}$. We
would like to extend this to the case of interest here involving
additional right-handed neutrinos and the $\Sigma$ field. However
as these are $SU(3)\otimes SU(2)_L\otimes U(1)$ singlets they do
not affect the anomaly structure discussed in \cite{IR}. If,
however, the gauge group is extended\footnote{This seems
inevitable if we are to explain the symmetric mass matrix
structure assumed here.} to $SU(3)\otimes
SU(2)_L\otimes SU(2)_R\otimes U(1)$ there will be constraints on
the
$\Sigma$ charge. From the left- right- symmetry  we may conclude
that the contributions to the $SU(2)_L^2 U(1)_{FD}$ and
$SU(2)_R^2 U(1)_{FD}$ anomalies are the same from the matter
fields. In \cite{IR} the Higgs fields had to be neutral under
$U(1)_{FD}$ to maintain anomaly cancellation (via the Green
Schwarz mechanism\cite{GSM}). The same consideration applied to
the $\Sigma$ field requires $Q_{\Sigma}+2a_1$ to be zero
corresponding to the cases 1 and 7. We note that these cases were
consistent with neutrino masses generating structure and solar
neutrino oscillations via the MSW effect.

%%%%%%%%%%%%%%%%%%%%

\end{document}